\documentclass[showpacs,preprint,aps,prb,twocolumnn,superscriptaddress,showpacs]{revtex4}

\usepackage{graphicx}
\usepackage{dcolumn}

\begin{document}

%\preprint{}

\title{Vibrational frequencies of light impurities in silicon}

%\author{J.M. Pruneda{$^{a}$}, S.K. Estreicher{$^{b}$}, J. Junquera{$^{c}$},
%J. Ferrer{$^{a}$}, P. Ordej\'on{$^{d}$}}

%\address{{$^{a}$}Departamento de F\'{\i}sica, Facultad de Ciencias, 
%Universidad de Oviedo, 33007 Oviedo, Spain.}
%\address{{$^{b}$}Department of Physics, Texas Tech University, 
%Lubbock, TX 7940 9-1051, USA}
%\address{{$^{c}$}Dep. de F\'{\i}sica de la Materia Condensada C-III, 
%Universidad Aut\'onoma, E-28049 Madrid, Spain}
%\address{{$^{d}$}Institut de Ci\`encia de Materials de Barcelona -- CSIC, 
%Campus de la UAB, E-08193 Bellaterra, Spain}

\author{J.M. Pruneda}
\affiliation{Departamento de F\'{\i}sica, Facultad de Ciencias, 
Universidad de Oviedo, C./ Calvo Sotelo s/n, 33007 Oviedo, Spain}
\email[Corresponding Author:]{pruneda@icmab.es}

\author{S.K. Estreicher}
\affiliation{Department of Physics, Texas Tech University, 
Lubbock, TX 7940 9-1051, USA}

\author{J. Junquera}
\affiliation{Dep. de F\'{\i}sica de la Materia Condensada C-III, 
Universidad Aut\'onoma, E-28049 Madrid, Spain}

\author{J. Ferrer}
\affiliation{Departamento de F\'{\i}sica, Facultad de Ciencias, Universidad de Oviedo, C./ Calvo Sotelo s/n, 33007 Oviedo, Spain.}

\author{P. Ordej\'on}
\affiliation{Institut de Ci\`encia de Materials de Barcelona -- CSIC, 
Campus de la UAB, E-08193 Bellaterra, Spain}


%\author{}
%\thanks{}


\date{\today}

\begin{abstract}
We have developed a formulation of density functional perturbation theory
for the calculation of vibrational frequencies in molecules and solids, 
which uses numerical atomic orbitals as a basis set for the electronic states.
The (harmonic) dynamical matrix is extracted directly from the first order 
change in the density matrix with respect to infinitesimal atomic 
displacements from the equilibrium configuration. 
We have applied this method to study the vibrational properties of a 
number of hydrogen-related complexes and light impurities in silicon. 
The diagonalization of the dynamical matrix provides the vibrational modes and
frequencies, including the local vibrational modes (LVMs) associated 
with the defects.  In addition to tests on simple molecules, results 
for interstitial hydrogen, hydrogen dimers, vacancy-hydrogen and 
self-interstitial-hydrogen complexes, the boron-hydrogen pair,
substitutional C, and several O-related defects in c-Si are presented. The
average error relative to experiment for the $\sim$
60 predicted LVMs is about 2\%  with most highly harmonic modes 
being extremely close and the more anharmonic ones within 5-6\% of the 
measured values.
\end{abstract}

%\pacs{{\it Keywords:} Density Functional Theory; 
%Local Vibrational Modes; Impurities in Silicon}
%\keywords{}

\maketitle


\section{Introduction}
The knowledge of the structures of impurities and defects is an essential
prerequisite for understanding the electrical and optical changes that these
complexes induce in semiconductors such as crystalline silicon\cite{Pe92,SKE95}.
The presence of light impurities such as H, B, C or O, results in the appearance
of infra-red (IR) or Raman active local vibrational modes (LVMs) usually well
isolated from the frequency range of the phonons of the host material. 
The observation of LVMs coupled with
isotope substitutions and uniaxial stress measurements provide precious
information about the type and number of impurity atoms involved and the
symmetry of the defect. However, these data are rarely sufficient to identify
unambiguously the defect. Since the early days of Stein\cite{Stein}, a large
number of vibrational modes have been identified through the interplay of
experiment and theory.

The calculation of LVMs at the {\it ab-initio} level provides a critical link
between theory and experiment.  This is particulary true in the case of 
hydrogen,
since it binds covalently in the immediate vicinity of many impurities and
defects thus giving rise to a number of LVMs in the range $\sim$ 800 to
$\sim$2200 cm$^{-1}$. Other common impurities in Si which produce LVMs are B, C,
O, but any element lighter than Si can in principle be observed by LVM
spectroscopy.

The computation of systematically accurate vibrational frequencies is a
challenge for first principles theory, given their sensitivity on details
of bonding geometry and electronic structure. Typical accuracies in 
the calculated vibrational modes for light impurities in silicon in 
former works are within 3-10\%\ of the experimental data, which means 
in some cases a deviation of over 100 cm$^{-1}$.

Various approaches have been used to calculate LVMs, from semiempirical
models\cite{semi}, to {\it ab-initio} Hartree-Fock\cite{HF} and density
functional theory\cite{VdW,Jones,ChCh}. In most cases, frequencies
are calculated in the spirit of the {\it frozen phonon} approximation. 
One computes the total energy of the system in the equilibrium configuration 
(that in which the forces acting on the atoms are zero) and then for small
displacements of selected atoms (either individually or in the direction 
of a normal mode, if this is known). 
%[The response to this displacement 
%of the other atoms in the cluster or cell is either neglected or partly 
%included by allowing the relaxation of some nearest neighbors.] 
The actual value of the atomic displacement (typically a few hundredths 
of an \AA) and the response of the nearby atoms are parameters chosen 
by the user.
One can either fit the energy vs. displacement to a polynomial and extract
a specific vibrational mode\cite{VdW,anharm}, or compute the dynamical 
matrix by finite differences\cite{Jones}. When a few specific modes is all
that is needed, only the movement of the atoms involved in those modes is
considered. In these methods, it is not possible to completely isolate the 
harmonic contributions from the anharmonic ones, since finite displacements
always involve some anharmonic effects. For this reason, the frequencies 
obtained in this approach are sometimes referred to as 
{\it quasi-harmonic}\cite{Jonesb}.

One can also calculate vibrational properties from constant-temperature
molecular-dynamics simulations, for instance by extracting selected 
frequencies from the velocity-velocity autocorrelation function\cite{v-v}
or by using more sofisticated spectral estimators, like the 
MUltiple SIgnal Classification (MUSIC) algorithm.\cite{music,kohanoff} 
This is computationally exhausting, since long molecular dynamics runs 
are required, but potentially very accurate.\cite{kohanoff} 
This also allows the calculation of frequencies as a function of temperature.

However, the calculation of vibrational frequencies does not neccesarily require
the actual displacement of the atoms, as in the methods described above.
Linear response theory (in particular through the application of perturbation theory
in density functional theory) has been thoroughly used in the past\cite{Baroni-rmp} 
to compute the response of the system to infinitesimal atomic displacements,
and from that, the vibrational frequencies in the harmonic approximation.
This can be done with the only knowledge of the electronic solution in the 
equilibrium configuration. The advantage of this approach is that anharmonic
effects are elliminated, and that no reference is needed to explicit finite atomic
displacements. Besides, this approach allows to compute phonons with arbitrary
{\bf q} vector in crystalline systems, without having to consider a supercell
conmensurate with the periodicity of the phonon, as it is required in the 
frozen phonon and molecular dynamics approaches. 

We propose a method, based on density functional perturbation
theory (DFPT), to compute vibrational frequencies in the harmonic approximation.
We use a basis set of numerical atomic orbitals to expand the electronic
wavefunctions, which makes the method computationally very efficient, and
allows us to calculate systems with a large number of atoms. 
We have applied it to make a systematic study of 
a number of defect centers in silicon, involving light impurities
and their complexes with intrinsic defects (vacancies and self-interstitials). 
In most cases, the comparison of the calculated and measured vibrational  
frequencies is very favorable, improving on the results obtained by other
approaches.

The outline of this paper is as follows.  We first discuss the theoretical
method and the model used to describe the defects. Then we compare the
vibrational properties obtained for a variety of complexes with experimental 
data as well as other first-principles calculations in the literature.
Finally, we discuss the results.

\section{Methodology}
\subsection{Ground State Description}

In this work, we use the fully self-consistent {\it ab-initio} code 
{\sc Siesta}\cite{siesta2, siesta3}. 
The electronic energy is obtained from density-functional 
theory (DFT)\cite{Kohn64,Kohn65} within the local density approximation.  
The exchange-correlation potential is that of Ceperley-Alder\cite{ca} as
parameterized by Perdew and Zunger.\cite{pz} Norm-conserving
pseudopotentials\cite{tm2} in the Kleinman-Bylander form\cite{kb} are used to
remove the core electrons from the calculations.  

The valence electron
wavefunctions are described with numerical linear combinations of
atomic orbitals (LCAO) of the Sankey type,\cite{sankey} but generalized   
to be arbitrarily complete with the inclusion of multiple-zeta orbitals  
and polarization states.\cite{siesta1} These orbitals are numerical 
solutions of the free atom with the appropriate pseudopotential, 
and are extrictly zero beyond some cutoff radius. This makes
the calculation of the Kohn-Sham hamiltonian to scale linearly
with the number of atoms,\cite{siesta2,siesta3} allowing calculations in very 
large systems with a modest computational cost.

In the present work, the basis sets include single-zeta (SZ), double-zeta
(DZ), double-zeta plus polarization (DZP), and triple-zeta plus polarization
(TZP). A DZP basis includes two sets of $s$ and $p$'s plus one set of $d$'s 
on Si, O, C or B, two $s$'s and one set of $p$'s on H. The radial cutoff 
of the atomic orbitals was determined as described in Ref. \onlinecite{siesta1}, 
with an energy shift of 0.5 Ry, and a split-norm of 0.15 for all the species
except H, for which the split-norm was 0.5.
The charge density is
projected on a real space grid with an equivalent cutoff of 90 to 150$\>Ry$
to calculate the exchange-correlation and Hartree potentials. The host
crystal is represented by a periodic supercell of 64 host atoms, and the
{\bf k}-point sampling is reduced to the $\Gamma$ point. This restriction
appears to be quite sufficient for the calculation of vibrational spectra.
Tests have been performed for selected defects in a 128 host atoms cell
and the results are within a few wavenumbers of those obtained in the
64 host atoms cell.

In order to determine the equilibrium structure of the defects studied,
we have relaxed all the atomic coordinates with a conjugate gradient     
algorithm, reaching a tolerance in the forces of F$_{max}<$0.01eV/\AA.  
The dynamical matrix for the whole cell is computed (see below) from
this ground state and its eigenfrequencies and eigenmodes obtained.

\subsection{Linear Response Theory}
A new implementation of DFPT has been developed to compute the electronic
response to {\it infinitesimal\/} atomic displacements.  As is well known
from the ``$2n+1$'' theorem in quantum mechanics,\cite{gonze} the first-order 
change of the electronic wavefunction in a perturbative expansion allows the 
computation of the second-order change in the energies.  This implies 
that only the properties of the unperturbed ground state are needed to 
obtain the linear response of the system.
The extension of this theorem to DFT is that the knowledge of the first-order
change in the electronic density determines variationally the second-order
change in the energy. In this way, we obtain analytically the dynamical matrix
from the gradient of the density relative to atomic displacement and, from it,
the vibrational properties, without physically displacing any atom. 

We describe briefly here the key points of our formulation. A complete report
will be published elsewhere.\cite{pruneda} 
The change in the electronic wavefunction is obtained 
by solving the first-order perturbation expansion of the Schr\"odinger equation
(Sternheimer equation\cite{Sternheimer})
\begin{equation}
\label{Sternheimer}
\delta\hat{H}\psi_{0,i} + \hat{H}_{0}\delta\psi_{i} =
\delta\epsilon_{i}\psi_{0,i} + \epsilon_{0,i}\delta\psi_{i}
\end{equation}
where $\psi_{0,i}$ are the ground state electronic wavefunctions and
$\delta\psi_{i}$ the first order perturbation of $\psi_i$ when an atom
is displaced (if we consider atom $\alpha$, this would be
$\partial_\alpha\psi_i$).  As we expand these wave functions in terms of
atomic orbitals:
\begin{equation}
\psi_{0,i}({\bf r}) = \sum_{\mu}c_{i\mu}\phi_\mu({\bf r}-{\bf R}_\mu) \\
\end{equation}
the derivatives can be directly written in terms of the  
derivatives of the atomic orbitals:
\begin{equation}
\partial_\alpha\psi_i({\bf r}) =
          \sum_{\mu}[\partial_\alpha c_{i\mu}\phi_\mu({\bf r}-{\bf R}_\mu) +
                    c_{i\mu}\partial_\alpha\phi_\mu({\bf r}-{\bf R}_\mu)]\>.
\end{equation}
Only the orbitals $\phi_\mu$ centered on atom $\alpha$ appear in the last term
which is equal to $-\nabla\phi_\mu({\bf r}-{\bf R}_\mu)$.  The change in
the coefficients, $\partial_\alpha c_{i\mu}$, is obtained from equation
\ref{Sternheimer}.  $\partial\psi_{i}$ is then used to compute the
perturbation in the electronic density
\begin{equation}
\partial_\alpha\rho(\bf{r})=
\sum_{i=1}^{occ}[\partial_\alpha\psi_i^*\psi_i+\psi_i^*\partial_\alpha\psi_ {i}]\>.
\end{equation}
This allows the computation of the dynamical matrix, by explicit derivation of
the forces on all the atoms $\beta$ in the system (the expressions of which
can be found in Refs. \onlinecite{siesta2} and \onlinecite{siesta3} for our
approach) with respect to the infinitesimal desplacement of one of them ($\alpha$):
\begin{equation}
(M_\alpha M_\beta)^{1/2} {\bf D}_{\alpha\beta}=
{{\partial^{2}{E}}\over{\partial{{\bf R}_{\alpha}}\partial{{\bf R}_{\beta}}}}
= \partial_\alpha {\bf F}_\beta
\end{equation}
%
%\begin{eqnarray}
%\nonumber
%E&=&\sum_{i=1}^{occ}\langle\psi_i|\hat{H}|\psi_i\rangle \\
%\nonumber
%\partial^2_{\alpha\beta} E &=&
%\sum_{i=1}^{occ}
%\{2{\rm Re}[\langle\partial_\alpha{\psi_i}|\hat{H_0}|\partial_\beta\psi_i\rangle
%] + \langle\psi_i|\partial_{\alpha\beta}^{2}\hat{H}|\psi_i\rangle   \\  
%&& + 
%2{\rm Re}[\langle\partial_\alpha{\psi_i}|\partial_\beta \hat{H} |\psi_i\rangle +
%\langle\partial_\beta{\psi_i}|\partial_\alpha \hat{H} |\psi_i\rangle] \}\>,
%\end{eqnarray}
We remark that only terms up to first order in the electronic wavefunctions
appear in the resulting formulas.
Note that only the linear effects are obtained in this method, which is
consistent with the harmonic approximation implicitly assumed in the
diagonalization of the dynamical matrix.  Thus, one expects to obtain
high-quality frequencies for the vibrational modes that are harmonic, but
the frequencies of modes involving large anharmonic contributions will be
less accurate.  
Although phonons with arbitrary {\bf q} vector can be obtained in
the pressent approach, here we only calculate vibrations which are
periodic with the simulation supercell ({\it i.e.}, ${\bf q}=\Gamma$),
since we are interested in LVMs.

\section{Results}

Tests of the method for free SiH$_{4}$, CO, CO$_2$ and H$_{2}$ lead to
very good agreement with experiment (see table \ref{molecules}).  Using an
appropriate basis set reveals to be essential to reproduce accurate
frequencies.  In most of the cases a DZ set gives good values, but in some
configurations a more complete basis is required. This is particularly true   
for bending modes.  Oxygen likes to have polarization orbitals, and thus the
frequencies are better when these are included. Note that the vibrational
frequencies are more sensitive to the basis set size than the structural
properties, such as bond lengths. In general, the largest improvements in
the frequencies correspond to going from SZ to DZ, then DZ to DZP, but TZ  
basis produce only marginal improvements.

\renewcommand{\tabcolsep}{0.2cm}
\begin{table*}[h]
%\begin{minipage}[t]{0.44\textwidth}
\caption{\label{molecules}
Calculated and measured\cite{nist} frequencies for free $\rm SiH_4$,
$\rm CO$, $\rm CO_2$ and $\rm H_2$ molecules with various basis sets.
%Other authors find also accurate frequencies for free
%$\rm SiH_4$ stretching modes ($\omega_T=2123(-3\%)$ and $\omega_A=2190(+0\%)$)
%although their ordering is incorrect.
}
\begin{center}
\begin{tabular}{|c|rrrrrr|r|}
\hline
        & \multicolumn{1}{c}{SZ}  &
          \multicolumn{1}{c}{SZP} &
          \multicolumn{1}{c}{DZ}  &
          \multicolumn{1}{c}{DZP} &
          \multicolumn{1}{c}{TZ}  &
          \multicolumn{1}{c}{TZP} & \multicolumn{1}{|c|}{Expt.} \\
\cline{2-8}
& \multicolumn{6}{c|}{$SiH_4$} & \\
$T_2$  & 2045 & 2064 & 2160 & 2153 & 2173 & 2159 & 2191 \\
$A_1$  & 1970 & 1974 & 2110 & 2116 & 2131 & 2125 & 2187 \\
 $E$   &  800 &  862 &  921 &  929 &  926 & 929  &  975 \\
$T_2$  &  701 &  772 &  806 &  818 &  817 & 821  &  914 \\
\hline
& \multicolumn{6}{c|}{$H_2$} & \\
 $A$   & 3619 & 3670 & 4194 & 4185 & 4193 & 4191 & 4161 \\
\hline
& \multicolumn{6}{c|}{$CO$} & \\
 $A$   & 1681 & 1823 & 1885 & 2088 & 1945 & 2183 & 2170 \\
\hline
& \multicolumn{6}{c|}{$CO_2$} & \\
$A_1$  & 2118 & 2355 & 2235 & 2277 & 2224 & 2394 & 2349 \\
$A_2$  & 1107 & 1216 & 1200 & 1241 & 1209 & 1331 & 1333 \\
 $E$   &  478 &  558 &  547 &  583 &  560 & 635  &  667 \\
\hline
\end{tabular}
\end{center}
%\end{minipage}
\end{table*}

 \renewcommand{\tabcolsep}{0.12cm}

A number of defects containing light impurities in c-Si are now considered.
These are H at a bond-center (BC) site, H dimers (H$_{2}$ and H$_{2}^{*}$),
the hydrogenated vacancy ($\rm VH_{n},n=1,2,3,4$) and the saturated divacancy
$\rm V_{2}H_{6}$, the self-interstitial-hydrogen IH$_{2}$ complex, the
$\rm\{B,H\}$
pair, substitutional C, interstitial O, and two charge states of the A-center
(oxygen-vacancy complex). These embrace a range of Si-X bonding
configurations, with X=H, C, O and B.  In most cases, we have tried different
basis sets in order to check, or improve the accuracy of our calculations
when the chemical properties are particularly complex.  We show frequencies
for the DZ and DZP basis sets. Although the diagonalization of the dynamical
matrix gives all the ($\Gamma$ point) modes in the cell, 
we present here only the stretching
and some bending LVMs. We also obtain the eigenvectors which we use to
determine the symmetry of the corresponding vibrational modes. Our results
are compared with available experimental data, and with theoretical 
frequencies obtained by other authors using first-principle DFT.

 \subsection{$\bf H_{BC}$, $\bf H_{2}$ and $\bf H_{2}^{*}$}

 At the BC site,\cite{SKEBCH} hydrogen exists in the +1 and the 0 charge states.
 In the latter case, the odd electron does not participate in the bonding but
 resides in a
 non-bonding orbital primarily localized on the two Si atoms adjacent to
 the proton. Chemically, this is a 3-center, 2-electron bond, very much
 like the type of H bonding occurring in boron hydrides.  The
 bond is somewhat compressed because optimal relaxation (no Si second
 nearest neighbors) would likely result in a longer Si-H-Si bond and a
 frequency lower than observed.  This suggests that as H moves along the trigonal
 axis, it tends to form something like Si...H-Si then Si-H-Si then Si-H...Si,
 a process which is highly anharmonic.
 The calculated frequency is in table \ref{HH}.

Raman\cite{H2Raman} and IR\cite{H2,H2IR} measurements of H$_2$ in silicon 
reveal a considerable softening of the stretching mode with respect to the
frequency of H$_2$ in the gas phase.
A number of calculations (for a review, see Ref. \onlinecite{SKEH2}) found the
molecule to be stable at the tetrahedral interstitial (T) site. The electron
affinity of the Si atoms surrounding H$_2$ is at least partly responsible for
a small charge transfer from H$_2$ to its Si neighbors, which results in a
weakening of the H-H bond. Even though the H-H stretch mode is not expected 
to be fully harmonic, our calculated frequency is close to the experimental one
(table \ref{HH}). Note that the errors relative to experiment in the D$_2$
and HD frequencies are very different than the error in the H$_2$ frequency.
This is also a clear feature of these frequencies when they are
calculated dynamically from the v-v autocorrelation function.\cite{skeicds}

\begin{figure}
\includegraphics{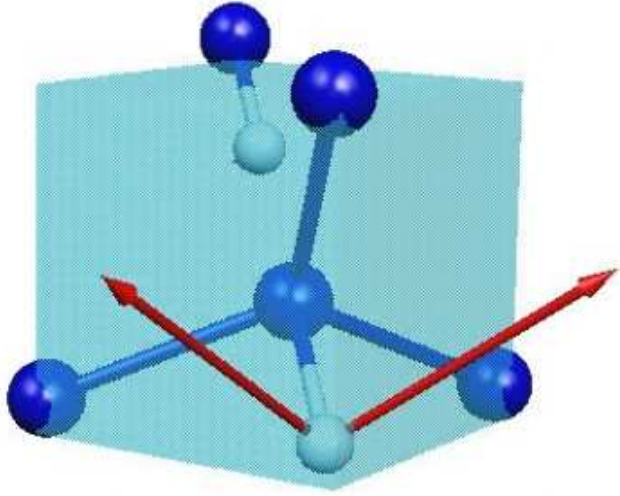}
\caption{\label{fig:wide}Calculated structure of $\rm H_2^{*}$ complex in silicon. 
Dark spheres are Si atoms, white spheres are H.  The two perpendicular arrows 
represent the H$_{AB}$ wag eigenmodes at 842 cm$^{-1}$.  The twofold degenerate 
wagging mode for  H$_{BC}$ is found at 560 cm$^{-1}$.}
\end{figure}

 The trigonal H$_2^*$ defect\cite{ChCh} consists of one hydrogen atom close to
 the antibonding (AB) site, and the other near the BC site (see figure 1).
 The two H atoms are inequivalent.  The Si-H$_{AB}$ bond length is slightly
 longer than the Si-H$_{BC}$ one (we obtain 1.580 and 1.510 \AA, respectively)
 which gives rise to different stretch frequencies for the two atoms: 2135 
 cm$^{-1}$ for H$_{BC}$ and 1750 cm$^{-1}$ for H$_{AB}$. 
 We also obtain two degenerate waggind modes, associated with the H$_{BC}$ atom,
 at 560 cm$^{-1}$, and two wagging modes with 839 and 843 cm$^{-1}$, 
 related to H$_{AB}$. The latter should be degenerate, but small inaccuracies
 in the atomic relaxations render them at slightly different frequencies.

 \begin{table*}[h!]
 %\begin{minipage}[h]{0.44\textwidth}
 \caption{\label{HH} Calculated and measured frequencies for $\rm H_{BC}$   
 (our calculation is spin polarized for $\rm H_{BC}^0$ and spin
 averaged for $\rm H_{BC}^+$), $\rm H_2$ in the $\langle100\rangle$ alignment,
 and $\rm H_2^*=H_{BC}H_{AB}$.  A DZ basis was used for all these complexes.
 The errors relative to experimental values are in parenthesis.
 ($a$) is Ref. \onlinecite{BCH},
 ($b$) is Ref. \onlinecite{JoBCH},
 ($c$) is Ref. \onlinecite{VdWBCH},
 ($d$) is Ref. \onlinecite{H2},
 ($e$) is Ref. \onlinecite{JoH2},
 ($f$) is Ref. \onlinecite{VdWH2},
 ($g$) is Ref. \onlinecite{JoH2+},
 ($h$) is Ref. \onlinecite{VdWH}, and
 ($i$) is Ref. \onlinecite{Kim}.
 Were $\rm H_2$ a classical dumbbell, its wag modes would be at 731
 and 860 cm$^{-1}$ (see discussion in Ref. \onlinecite{SKEH2}).
 }
 \begin{center}
 \begin{tabular}{|crr@{}lr@{}lr@{}lr@{}l|}
 \hline
  &  \multicolumn{1}{c}{expt.}         &
     \multicolumn{2}{c}{\bf this work}      &
     \multicolumn{6}{c|}{other authors} \\
 \hline
  \multicolumn{10}{|c|}{$\rm H_{BC}$} \\
   $H_{BC}^{+}$ &  1998$^{a}$ &
                   {\bf 1891}  &($\bf -5\%$) &
                   2203$^{b}$ & ($+10\%$) & 2210$^{c}$ & ($+11\%$) & & \\
   $H_{BC}^{0}$ &  \multicolumn{1}{c}{$-$} &
                   {\bf 1813}  &  & 1768$^{b}$ & & 1945$^{c}$ & & &\\
 \hline
 \multicolumn{10}{|c|}{$\rm H_{2}$ in Si} \\
     $H_2$     &  3618$^{d}$ &
                   {\bf 3549}  & ($\bf -2\%$) & 3607$^{e}$ & ($+0\%$) &
                                                  3396$^{f}$ & ($-6\%$) &
                                                  3260$^{i}$ & ($-9\%$) \\
       $DH$      &  3265$^{d}$ &
                     {\bf 3081}  & ($\bf -6\%$) & 3129$^{e}$ & ($-4\%$) &
                     &   &      & \\
     $D_2$     &  2643$^{d}$ &
                     {\bf 2511}  & ($\bf -5\%$) & 2559$^{e}$ & ($-3\%$) &     &
 &
  & \\
 \hline
                 &  2062$^{g}$ &  {\bf 2135}  & ($\bf +3\%$) & 2164$^{g}$ & ($+5
\%$)
 & 2100$^{h}$ & ($+2\%$) &
  1945$^{i}$ & ($-5\%$) \\
    $H_{2}^{*}$  &  1838$^{g}$ &  {\bf 1750}  & ($\bf -5\%$) & 1844$^{g}$ & ($+0
\%$)
  & 1500$^{h}$ & ($+18\%$) &
  1677$^{i}$ & ($-9\%$)\\
                 & 817$^{g}$  & {\bf 843/839} & ($\bf +3\%$) & 1002$^{g}$ &
 ($+22\%$)  &  &  & 711$^{i}$ & ($-13\%$) \\
 \hline
 \end{tabular}
 \end{center}
 %\end{minipage}
 \end{table*}

 \subsection{Hydrogen and native defects}
 A considerable number of IR and Raman lines are related to H--intrinsic
 defect complexes.  It has been noted\cite{V-I,V2H6} that vibrational
 modes above 2000 cm$^{-1}$ are
 mainly related to H in vacancies, while those lines below
 2000 cm$^{-1}$ are predominant for the H-self-interstitial systems or H
 at AB sites.  A large number of geometrical configurations may
 lead to very similar vibrational lines, making it difficult to
 identify these defects unambiguously.  As noted by other groups
 \cite{SKE95,JoV,Singh,Deak}, the streching frequencies in VH$_n$
 ($n=1,2,3,4$) systems increases with $n$ due to the repulsive H-H
 interaction.  Thus, the highest IR line is that of VH$_4$.  The
 Si-H bonds point toward the center of the vacancy along the 
 trigonal axes.   

 In our calculations (table \ref{VH}), VH has monoclinic symmetry, and the H
 oscillates  parallel to the $\langle111\rangle$ direction.
 In the orthorhombic VH$_{2}$, the two
 equivalent H have stretching modes along the $\langle100\rangle$ and
 $\langle001\rangle$ directions.  The frequencies for these modes are
 2121 and 2144 cm$^{-1}$ respectively.  VH$_{3}$ has $C_{3v}$
 symmetry.  The $A$ singlet involves the movement of the three H atoms towards
 the vacancy, while in the twofold degenerate $E$ mode one of the atoms
 moves in opposition. VH$_{4}$ has $T_{d}$ symmetry.  In addition to the
 threefold degenerate $T_{2}$ mode at 2205, we obtain an IR-inactive
 singlet $A_{1}$ mode at 2265 cm$^{-1}$.

 The vibrational modes of $\rm V_{2}H_{6}$ are almost identical
 to those of VH$_{3}$: The fully saturated divacancy behaves very much
 like two weakly coupled VH$_{3}$ complexes. The $A_{2}$ singlet at 2176 cm$^{-1}$
 induces a dipole along the $\langle111\rangle$ direction.  In addition to
 this mode and the IR-active $E$ doublet, we obtain two IR-inactive modes
 at 2186 cm$^{-1}$ and 2134 cm$^{-1}$.

 \begin{table*}[t]
 %\begin{minipage}[h]{0.44\textwidth}
 \caption{\label{VH} Calculated and measured frequencies for
 stretching modes in $\rm VH_n$ ($n=1,2,3,4$) and
 $\rm V_2H_6$.
 ($a$) is Ref. \onlinecite{V2H6},
 ($b$) is Ref. \onlinecite{JoV}, and
 ($c$) is Ref. \onlinecite{Lavr}.
 The error relative to experiment is in parenthesis.  Our
 frequencies were obtained with a DZP basis.  The measured values for   
the VH$_2$D, VHD$_2$ and VD$_3$ as published in Ref. \onlinecite{V2H6} are now  
believed to belong to a divacancy complex\cite{BBpriv} and are therefore
not listed here.}
 \begin{center}
 \begin{tabular}{|cccl|cccc|} \hline
 & expt.$^{a}$ & \multicolumn{1}{c}{{\bf this work}} &
 \multicolumn{1}{c|}{Ref.$^{b}$} &
  & Expt.$^{a}$ & \multicolumn{1}{c}{{\bf this work}} &
 \multicolumn{1}{c|}{Ref.$^{b}$}\\
 \hline
      \multicolumn{4}{|c|}{VH} & \multicolumn{4}{c|}{$\rm VH_{4}$} \\
 $A'$ & 2038 & {\bf 1971($\bf -3\%$)} &  2248($+10\%$) &
 $A_{1}$ & 2257$^{c}$ & {\bf 2265($\bf +0\%$)} & 2404($+6\%$) \\
      \multicolumn{4}{|c|}{$\rm VD$} &
 $T_{2}$ & 2222 & {\bf 2205($\bf -1\%$)} & 2319($+4\%$)  \\
 $A'$ & 1507 & {\bf 1418($\bf -6\%$)} & 1613($+7\%$)  &
 \multicolumn{4}{c|}{$V\rm H_{3}D$} \\
 \cline{1-4}
 \multicolumn{4}{|c|}{$\rm VH_{2}$} &
 $A_{1}$ & 2250 & {\bf 2251($\bf +0\%$)} & 2384($+6\%$) \\
 $A_{1}$ & 2144 & {\bf 2163($\bf +1\%$)} & 2316($+7\%$) &
 $E$     & 2224 & {\bf 2205($\bf -1\%$)} & 2319($+4\%$) \\
 $B_{1}$ & 2121 & {\bf 2132($\bf +1\%$)} & 2267($+7\%$) &
 $A_{1}$ & 1620 & {\bf 1594($\bf -2\%$)} & 1677($+3\%$) \\
 \multicolumn{4}{|c|}{$\rm VHD$} & \multicolumn{4}{c|}{$\rm VH_{2}D_{2}$}\\
 $A'$    & 2134 & {\bf 2135($\bf -0\%$)} & 2292($+7\%$) &
 $A_{1}$ & 2244 & {\bf 2235($\bf -0\%$)} & 2364($+5\%$) \\
 $A'$    & 1555 & {\bf 1551($\bf -0\%$)} & 1641($+5\%$) &
 $B_{1}$ & 2225 & {\bf 2204($\bf -1\%$)} & 2319($+4\%$) \\
 \multicolumn{4}{|c|}{$\rm VD_{2}$} &
 $A_{1}$ & 1628 & {\bf 1603($\bf -1\%$)} & 1690($+4\%$) \\
 $A_{1}$ & 1564 & {\bf 1552($\bf -1\%$)} & 1658($+6\%$) &
 $B_{2}$ & 1615 & {\bf 1585($\bf -2\%$)} & 1663($+3\%$) \\
 $B_{1}$ & 1547 & {\bf 1532($\bf -0\%$)} & 1625($+5\%$) &
 \multicolumn{4}{c|}{$\rm VHD_{3}$} \\
 \cline{1-4}
 \multicolumn{4}{|c|}{$\rm VH_{3}$} &
 $A_1$   & 2236 & {\bf 2221($\bf -1\%$)} & 2342($+5\%$) \\
 $A_{1}$ & 2185 & {\bf 2158($\bf -1\%$)} & 2318($+6\%$) &
 $A_1$   & 1636 & {\bf 1613($\bf -1\%$)} & 1705($+4\%$) \\
 $E$     & 2155 & {\bf 2100($\bf -2\%$)} & 2256($+5\%$) &
 $E$     & 1616 & {\bf 1584($\bf -2\%$)} & 1664($+4\%$) \\
 \multicolumn{4}{|c|}{$\rm VH_{2}D$} &
 \multicolumn{4}{c|}{$\rm VD_{4}$} \\
%$A'$    & 2185 & {\bf 2140($\bf -2\%$)} & 2298($+5\%$) &
 $A'$    &      & {\bf 2140}             & 2298         &
 $A_1$   & no-IR & \multicolumn{1}{c}{{\bf 1623}} & \multicolumn{1}{c|}{1721} \\
%$A''$   & 2167 & {\bf 2101($\bf -3\%$}  & 2256($+4\%$)  &
 $A''$   &      & {\bf 2101}             & 2256          &
 $T_2$   & 1617 & {\bf 1584($\bf -2\%$)} & 1664($+3\%$)  \\
%$A'$    & 1580 & {\bf 1520($\bf -4\%$)} & 1632($+3\%$) &
 $A'$    &      & {\bf 1520}             & 1632         & 
 \multicolumn{4}{c|}{} \\
 \cline{5-8}
 \multicolumn{4}{|c|}{$\rm VHD_{2}$} &
 \multicolumn{4}{c|}{$\rm V_{2}H_{6}$} \\
%$A'$    & 2177  & {\bf 2121($\bf -3\%$)} & 2277($+5\%$) &
 $A'$    &       & {\bf 2121}             & 2277         &
 $A_{1}$ & 2190$^{c}$  & {\bf 2186($\bf -0\%$)} & \multicolumn{1}{c|}{$-$}  \\
%$A'$    & 1588  & {\bf 1534($\bf -3\%$)} &  1646($+4\%$) &
 $A'$    &       & {\bf 1534}             &  1646         &
 $A_{2}$ & 2191  & {\bf 2176($\bf -0\%$)} & \multicolumn{1}{c|}{$-$} \\
%$A''$   & 1575 & {\bf 1509($\bf -4\%$)} & 1619($+3\%$) & 
 $A''$   &      & {\bf 1509}             & 1619         &
 $E$     & 2166$^{c}$ & {\bf 2143($\bf -1\%$)} &  \multicolumn{1}{c|}{$-$}  \\
 \multicolumn{4}{|c|}{$\rm VD_{3}$} &
 $E$     & 2165 & {\bf 2134($\bf -1\%$)} &  \multicolumn{1}{c|}{$-$}  \\
%$A_{1}$& 1594 & {\bf 1547($\bf -3\%$)} & 1661($+4\%$) &
 $A_{1}$&      & {\bf 1547}             & 1661         &
 \multicolumn{4}{c|}{} \\
%$E$& 1576 & {\bf 1510($\bf -4\%$)} &  1619($+3\%$) &
 $E$&      & {\bf 1510}             &  1619         &
 \multicolumn{4}{c|}{} \\
 \hline
 \end{tabular}
 \end{center}
 %\end{minipage}
 \end{table*}

The IH$_2$ complex\cite{JoIH} has two equivalent and weakly coupled
hydrogen atoms, which yields two very similar stretching frequencies. 
Uniaxial stress measurements show that the two hydrogen atoms are equivalent.
Our relaxed structure has almost $C_{2v}$ symmetry, with the $A$ mode higher 
than the $B$ mode, confirming early results\cite{JoIH}. The deviation from 
perfect symmetry is due to the finite tolerance in the geometry optimization.
This deviation is seen when comparing the IHD and IDH complexes: they should 
be identical but we find their frequencies to be off by 2 cm$^{-1}$. 
Note that we reproduce the correct ordering for the bending modes of
IH$_2$ at 737 and 732 cm$^{-1}$ (table \ref{SiH}).

 \begin{table*}[t]
 %\begin{minipage}[t]{.44\textwidth}
 \caption{\label{SiH} Calculated frequencies with a DZ basis set compared
 with experimental and other theoretical results for $\rm IH_2$.
 ($a$) is Ref. \onlinecite{JoIH} and
 ($b$) is Ref. \onlinecite{JoV}.
 The error relative to experiment is in parenthesis.
 }
 \begin{center}
 \begin{tabular}{|crr@{}lr@{}lr@{}l|rr@{}lr@{}lr@{}l|}   
 \hline
  & expt.$^{a}$ & \multicolumn{2}{c}{{\bf this work}} &
    \multicolumn{4}{c|}{other authors}
  & expt.$^{a}$ & \multicolumn{2}{c}{{\bf this work}} &
    \multicolumn{4}{c|}{other authors}\\
 \hline
 \multicolumn{8}{|c|}{$\rm IH_{2}$} & \multicolumn{7}{c|}{$\rm ID_{2}$} \\ 
  $A$ & 1989 & {\bf 2007} &($\bf +1\%$) & 2107$^{b}$ & ($+6\%$) & 2145$^{a}$ & (
$+8\%$) &
  1448 & {\bf 1440} & ($\bf -1\%$) & 1510$^{b}$ & ($+4\%$) & 1540$^{a}$ & ($+6\%
$) \\
  %    &      &         & 1915$^{c}$ & ($+3\%$)   &      \\
  $B$ & 1986 & {\bf 2004} &($\bf +1\%$) & 2106$^{b}$ & ($+6\%$) & 2143$^{a}$ & (
$+8\%$) &
  1446 & {\bf 1438} & ($\bf -1\%$) & 1508$^{b}$ & ($+4\%$) & 1539$^{a}$ & ($+6\%
$) \\
  $B$ &   748 &  {\bf 737} &($\bf -1\%$) & & & ($A$) 775$^{a}$&($+3\%$)   &
       &  {\bf 609} &          &  & & 590$^{a}$            &         \\
  $A$ &  743 & {\bf 733} &($\bf -1\%$) & & & ($B$) 768$^{a}$&($+3\%$)   &
       &  {\bf 601} &          &  & &  583$^{a}$ &          \\
  $A$ &      & {\bf 716} &         & & &  ($B$) 736$^{a}$&          &
       &  {\bf 566} &          &   & & 564$^{a}$          &          \\
  $B$ &      & {\bf 711} &         & & & ($A$) 717$^{a}$&           &
       &  {\bf 562} &          &  & & 555$^{a}$           &          \\
 \hline
 \multicolumn{15}{|c|}{$\rm IHD/IDH$} \\
 \multicolumn{3}{|c}{
 \begin{tabular}{r}
 expt.$^{a}$ \\
 1988 \\ 
 1447 \\
  746 \\
      \\
 \end{tabular}
  } &
 \multicolumn{4}{c}{
 \begin{tabular}{r@{}l}
 \multicolumn{2}{c}{{\bf this work}} \\
  {\bf 2005/2007} & ($\bf +1\%$) \\
  {\bf 1440/1438} & ($\bf -1\%$) \\
   {\bf 733/736}  & ($\bf -2\%$) \\
   {\bf 714/714}  &          \\
 \end{tabular}
  } &
 \multicolumn{8}{c|}{
 \begin{tabular}{r@{}lr@{}l}
 \multicolumn{4}{c}{other authors}\\
  2106$^{b}$ & ($+6\%$) & 2144$^{a}$ & ($+8\%$) \\
  1509$^{b}$ & ($+3\%$) & 1540$^{a}$ & ($+6\%$) \\
             &          & 771$^{a}$  & ($+3\%$) \\
             &          & 727$^{a}$  &          \\
  \end{tabular}
  } \\
 \hline 
 \end{tabular}
 \end{center}
 %\end{minipage}
 \end{table*}

 \subsection{Oxygen,Carbon and Boron in Silicon}
Oxygen is a well-known impurity which is especially important in
Czochralski-grown Si, and a considerable amount of effort was focused in
understanding its properties.\cite{NATO}
We have computed the LVM frequencies for interstitial oxygen (O$_i$) and
two charge states of the vacancy-oxygen complex (A-center). The results are
in table \ref{O-Si}.  Frequencies for O$_i$ were computed with the oxygen 
placed at the BC site, where the probability of finding this delocalized 
atom is maximun, and the classical harmonic potential can better describe the
local modes.  The IR-active $A_{2u}$ mode corresponds to the 
asymmetric-stretching mode, while the $A_{1g}$ is the symmetric one.  
$E_u$ mode involves the movement of nearest silicon atoms with no 
participation of the oxygen\cite{Artacho}.
Finally, Table\ref{C-Si} shows the triply degenerate mode of substitutional
carbon as well as the LVMs associated with the $\rm\{B,H\}$ complex.\cite{SKE95}
 \begin{table*}[t]
 %\begin{minipage}[t]{.44\textwidth}
 \caption{\label{O-Si} Calculated and measured LVMs for interstitial O (O$_i$)
 and two charge states of the A-center (VO$^{0}$ and VO$^{-}$).
 ($a$) is Ref. \onlinecite{Kaiser} for O$_i$ 
and Ref. \onlinecite{VO} for VO$^{(0/-)}$,
 ($b$) is Ref. \onlinecite{Couti}, and
 ($c$) is Ref. \onlinecite{Pesol}.  For VO$^{(0/-)}$ a DZP basis was used for O
 and its Si nearest-neighbors, and a DZ basis for the other Si atoms.
 For O$_i$, a DZP basis was used for all the atoms in the cell.
 }
 \begin{center}
 \begin{tabular}{|c|crr@{}lr@{}lr@{}lr@{}l|}
 \hline
  && expt.$^{a}$ &  \multicolumn{2}{c}{\bf this work} &
    \multicolumn{4}{c}{Calc$^{b}$} & \multicolumn{2}{c|}{Calc$^{c}$} \\  
 \hline
  &\multicolumn{2}{c}{} & \multicolumn{2}{c}{$D_{3d}$} &
 \multicolumn{2}{c}{$D_{3d}$} &
 \multicolumn{2}{c}{$C_2$} & \multicolumn{2}{c|}{} \\
  &$A_{2u}$ & 1136 & {\bf 1131} & ($\bf -0\%$) & 1184 & ($+4\%$) &
                                      1108 & ($-2\%$) &
                                      1098      & ($-3\%$) \\
  O$_i$ &$A_{1g}$ &  618 & {\bf 607}  & ($\bf -2\%$) &  619 & ($+0\%$) &
                                       621 & ($+0\%$) &
                                       630      & ($+2\%$) \\  
  &$E_u$    &  518 & {\bf 538}  & ($\bf +4\%$) &  519 & ($+0\%$) &
                                       518 & ($+0\%$) &
           \multicolumn{2}{c|}{} \\
 \hline
 VO$^{0}$&$B_1$    &  836 & {\bf 861} &($\bf +3\%$) &
 \multicolumn{4}{c}{
 \begin{tabular}{r@{}l}
 839$^{a}$ & ($+0\%$)
 \end{tabular}} &            843 & ($+1\%$)  \\
 &$A_1$    &  534 & {\bf 546} &($\bf +2\%$) &
 \multicolumn{4}{c}{
 \begin{tabular}{r@{}l}
 548$^{a}$ & ($+3\%$)
 \end{tabular}}  & 540 & ($+1\%$)  \\
 VO$^{-}$&$B_1$    &  885 & {\bf 897} &($\bf +1\%$) &
 \multicolumn{4}{c}{
 \begin{tabular}{r@{}l}
 872$^{a}$ & ($-1\%$)
 \end{tabular}}  & 850 & ($-4\%$)   \\
 &$A_1$    &  545 & {\bf 558} &($\bf +2\%$) &
 \multicolumn{4}{c}{
 \begin{tabular}{r@{}l}
  532$^{a}$ & ($-2\%$)
 \end{tabular}}   & 539 & ($-1\%$)   \\
 \hline
 \end{tabular}
 \end{center}
 %\end{minipage}
 \end{table*}

\begin{table*}[t]
%\begin{minipage}[t]{.44\textwidth}
\caption{\label{C-Si} Calculated and measured frequencies for substitutional 
C and the $\rm\{B,H\}$ pair in Si.
($a$) is Ref. \onlinecite{Csexp},
($b$) is Ref. \onlinecite{Leary},
($c$) is Ref. \onlinecite{BHexp},
($d$) is Ref. \onlinecite{vWBH},
($e$) is Ref. \onlinecite{DeLeo}, and
($f$) is Ref. \onlinecite{BHexp2}.
A DZP basis was used for C and  its Si nearest-neighbors, and DZ for the 
other atoms.  For the $\rm\{B,H\}$ pair, a DZ basis was used for all the 
atoms.}
\begin{center}
\begin{tabular}{|c|crr@{}lr@{}l|}
\hline
& & expt. &  \multicolumn{2}{c}{\bf this work} &
  \multicolumn{2}{c|}{other authors} \\
\hline
C$_s$ & $T_d$    & 607$^{a}$ & {\bf 631} &($\bf +4\%$) & 662$^{b}$ & ($+9\%$)
\\
\cline{2-7}
    & $A$    &  1903$^{c}$ & {\bf 1958} &($\bf +3\%$) & 1830$^{d}$ & ($-4\%$) \\
$\rm\{B,H\}$ &        &             &      &         & 1880$^{e}$ & ($-1\%$) \\ 
    &  E     & 652    & {\bf 695}  &($\bf +6\%$) &        &
\\
\hline
\end{tabular}
\end{center}
%\end{minipage}
\end{table*}

\section{Conclusions}

We have presented a new development of DFPT, using localized atomic
wavefunctions as a basis set, and applied it to the study of LVMs
for light impurities in silicon.  In contrast
to other methods, the dynamical matrix is computed analytically without
actually displacing any atom from its equilibrium position.  The
calculations are based on the ground state density matrix as computed
with the {\sc Siesta} package.

Tests of the method for free molecules (SiH$_{4}$, H$_2$, CO, and CO$_2$)
show that this approach is highly accurate in situations where the
anharmonic contributions are small.  Note that the frequencies are
obtained at $T=0K$ while experimental data are obtained at low, but
non-zero, temperatures.

We have used a variety of basis set sizes to describe the electronic
wavefunctions.  In most cases, a DZ basis is quite sufficient to obtain
accurate atomic structures and vibrational frequencies. Larger basis sets
such as DZP improve the frequencies in situations that involve more
complex chemical bondings.
The defects included here are $\rm H_{BC}$, $\rm H_2$, $\rm H_2^*$,
$\rm VH_n$ (with n=1,2,3,4), $\rm V_2H_6$, $\rm IH_2$, the $\rm\{B,H\}$ pair,
substitutional C, interstitial O and two charge states of the A center.
These defects involve a wide range of bonding configurations.

The average error of the ~60 calculated modes relative to experiment is
about 2\%. In situations where large anharmonic contributions are present, the
accuracy of the method decreases somewhat ($5-6\%$).
This occurs, for example, when H is close to a BC position.
However, in most cases the calculated frequencies are in remarkable
agreement (0 to 2\%) with experimental data, implying that
this perturbative approach is totally justified and that the
ground state density matrix calculated with {\sc Siesta} is very reliable.

\section{Acknowledgments}
This work was supported by the spanish DGESIC (Project PB96-0080-C02). 
J.M.P. acknowledge F.P.I.  Grant from the Spanish Ministry of Science 
and Technology. 
S.K.E.'s research is supported in part by a grant from the
R.A. Welch Foundation, a contract from the National Renewable Energy
Laboratory, and a research award from the Humboldt Foundation.
P.O. acknowledges support from Fundaci\'on Ram\'on Areces and
Spain's MCyT project BFM2000-1312-C02-01, as well as the use
of computational resources from CESCA and CEPBA.

\end{document}